# Flowfield prediction of airfoil off-design conditions based on a modified variational autoencoder

Yunjia Yang, Runze Li, Yufei Zhang, Haixin Chen*[1]

*School of Aerospace Engineering, Tsinghua University, People's Republic of China*

**Abstract:** Airfoil aerodynamic optimization based on single-point design may lead to poor off-design behaviors. Multipoint optimization that considers the off-design flow conditions is usually applied to improve the robustness and expand the flight envelope. Many deep learning models have been utilized for the rapid prediction or reconstruction of flowfields. However, the flowfield reconstruction accuracy may be insufficient for cruise efficiency optimization, and the model generalization ability is also questionable when facing airfoils different from the airfoils with which the model has been trained. Because a computational fluid dynamic evaluation of the cruise condition is usually necessary and affordable in industrial design, a novel deep learning framework is proposed to utilize the cruise flowfield as a prior reference for the off-design condition prediction. A prior variational autoencoder is developed to extract features from the cruise flowfield and to generate new flowfields under other free stream conditions. Physical-based loss functions based on aerodynamic force and conservation of mass are derived to minimize the prediction error of the flowfield reconstruction. The results demonstrate that the proposed model can reduce the prediction error on test airfoils by 30% compared to traditional models. The physical-based loss function can further reduce the prediction error by 4%. The proposed model illustrates a better balance of the time cost and the fidelity requirements of evaluation for cruise and off-design conditions, which makes the model more feasible for industrial applications.

---

* Corresponding author: chenhaixin@tsinghua.edu.cn

**Nomenclature:**

*Symbols:*

| | | |
|---|---|---|
| $AoA$ | = | Angle of attack |
| $C_L$ | = | Lift coefficient |
| $C_D$ | = | Drag coefficient |
| $C_p$ | = | Pressure coefficient |
| $Ma$ | = | Freestream Mach number |
| $Re$ | = | Freestream Reynolds number |
| $(t/c)_{max}$ | = | Maximum relative thickness of an airfoil |
| $p$ | = | Pressure |
| $T$ | = | Temperature |
| $u$ | = | Velocity at $x$-direction |
| $v$ | = | Velocity at $y$-direction |
| $\mathbb{E}$ | = | Expectation |
| $c$ | = | Flow condition |
| $r$ | = | Reference cruise flowfield |
| $x$ | = | A sample of flowfield |
| $z$ | = | Latent code |
| $\mu$ | = | Mean of the Gaussian distribution |
| $\sigma$ | = | Standard deviation of the Gaussian distribution |

*Subscript / Superscript:*

| | | |
|---|---|---|
| $\sim$ | = | Dimensional value |
| $\wedge$ | = | Value given by estimation |
| $\infty$ | = | Freestream |

# I. Introduction

Decades of engineering have led to advances in the methods used to design civil aircraft in the pursuit of better transonic flight performance. In the early ages of aerodynamic design, designers mostly relied on their experience and understanding of aerodynamics. With the development of computational fluid dynamics (CFD) and optimization methods, aerodynamic optimization design has been widely used in industrial applications. However, optimizing the cruise efficiency based on a single-point design usually affects the off-design performance [1]-[4]. For example, increasing the lift-to-drag ratio under cruise conditions may lead to a thinner leading edge, which deteriorates performance at low speeds[4].

Multipoint optimization[1]-[7] is often applied to balance the performance factors under cruise and off-design conditions. Indicators of the flight envelope, such as the maximum lift coefficient[3] and the buffet onset[7], are also considered design constraints during optimization. Multipoint optimization requires additional CFD simulations under off-design conditions, which increases the computational cost. Consequently, many studies have tried to quickly predict off-design performance with empirical relations. For example, Korn's equation uses the airfoil lift coefficient and relative thickness to evaluate the drag divergence Mach number[8]. However, the empirical relation has limited accuracy and cannot provide off-design flowfields for further analysis.

In recent years, the success of machine learning (ML) in computer science has attracted wide attention from researchers. Many associated attempts have been made in the field of aerodynamic design[9], including fast prediction of airfoil performance[10]-[13] or its gradients[14] as a surrogate model, analyzing the correlation between optimization objectives and airfoil design variables[15],[16], and building reduced-order models (ROMs) for flowfields[17]. However, these models still rely on CFD to evaluate the airfoil performance during optimization. With the development of deep neural networks (DNNs), ML models have been proposed to directly reconstruct multicondition flowfields[18]-[24] in place of CFD to accelerate the evaluation process.

Two challenges stand in the way when these flowfield reconstruction models are applied to replace CFD in aerodynamic optimization. First, the flowfields generated by the DNN models have considerable errors. For example, the model proposed in [23] has a 5% error in the drag coefficient, which is even larger than the drag reduction ratio in one round of optimization. Second, the generalization of these models remains a major challenge for predictions throughout different freestream conditions and geometries. Since it is very expensive or even impossible to prepare a complete sample set that covers all freestream conditions and practical geometries, most of those elaborate models may act poorly when dealing with alien airfoil shapes, which may limit the search range in the optimization process.

In this paper, a novel DNN framework is proposed for multipoint aerodynamic predictions aimed at industrial applications to tackle these challenges. Instead of predicting flowfields for all freestream conditions, the proposed model only predicts off-design flowfields, while the cruise flowfield is evaluated by CFD and used as a prior reference. A modified variational autoencoder (VAE) model, called pVAE, is developed to extract the features from the cruise flowfield for predicting the off-design conditions. The p in the model name stands for the prior reference flowfield.

The pVAE model is developed because of the different fidelity requirements of the cruise and off-design points in aircraft design. The cruise flowfield and aerodynamic force coefficients are of great importance in the design process. The drag coefficient or the lift-to-drag ratio of a wing or an airfoil under the cruise condition, which is the primary design objective, requires high fidelity to push forward the design process. The off-design conditions are usually considered based on engineering constraints, so less accurate results predicted by a fast DNN model can be acceptable in the optimization process to save computational cost. Another motivation of the pVAE model is its better generalization ability. Instead of blindly facing any new compositions of parameters, the new model only needs to predict the evolution of the flowfields between the cruise point and other freestream conditions. Such an evolution relation might be similar for airfoils both inside and outside the training dataset. Consequently, compared to other models that predict flowfields of different freestream conditions separately, the present model with a reference cruise field can better capture the relationship of flowfields along with different freestream conditions.

In addition to the modification of the pVAE model's framework, physical-based loss functions are also discussed in this paper. Most of the flowfield prediction models in the literature[18]-[24] are inspired by the achievements of computer vision; however, there is a huge difference between a flowfield and an image because the flowfield explicitly follows physical laws, such as conservation laws of mass, momentum, and energy. Thus, several attempts have been made to apply physical laws to the machine learning process, such as the physics-informed neural networks (PINN) [25], or by adding loss terms derived from Navier–Stokes equations' residuals of predicted flowfields[21],[26]. In this paper, a new loss term is proposed from the demand for airfoil optimization, where the lift and drag coefficient is of great concern. The error between the aerodynamic coefficient integrated from the predicted flowfield and ground truth is added to the loss function in the training stage to increase the model prediction accuracy. The loss term based on the conservation of mass is also studied in this paper.

## II. Database establishment

In an industrial application, off-design points must be evaluated in the design process to avoid detriment to the overall performance. The freestream condition or the operating condition of an airfoil is determined by three variables: the freestream Mach number (*Ma*), the Reynold number (*Re*), and the angle of attack (*AoA*). The required lift coefficient ($C_L$) is determined by the aircraft weight, so $C_L$ is usually used to represent the operating condition in place of *AoA*, and a designer must adjust the *AoA* to obtain the required lift coefficient.

During a multipoint optimization process, flowfields under several flow conditions around the cruise point are evaluated to obtain the off-design performance[4]-[6]. Fig 1 depicts one possible selection of the combinations of $C_L$ and *Ma*, where eight off-design points are selected around the cruise point. Along the ordinate, the evolution of lift and drag performance with respect to the angle of attack is computed to evaluate the buffet characteristics and the drag performance of an airfoil. Meanwhile, along the abscissa, the variation of drag versus Mach number is also important to judge the drag creep phenomenon and drag divergence of an airfoil.

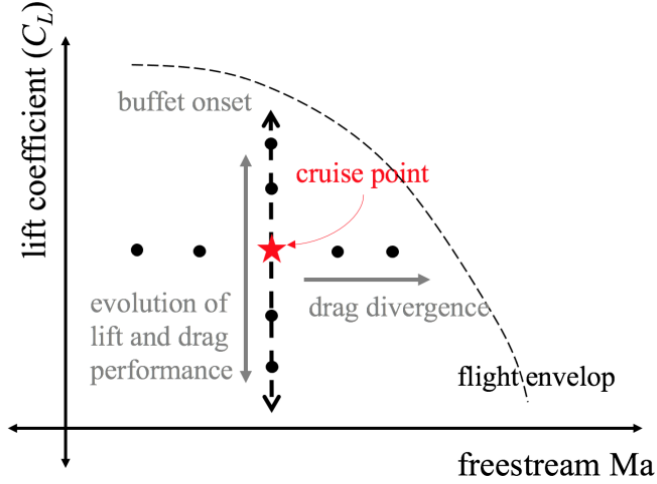

**Fig 1.** **A possible selection of operating conditions of the cruise and off-design points in a $C_L$-Ma plot**

(Each dot represents a combination of two parameters)

In this study, the off-design points of airfoils only under different lift coefficients are tested using the pVAE model to simplify the problem. The Mach number and Reynold number are set to 0.76 and $5.0 \times 10^6$, respectively. In this section, a database of flowfields is established to train the model.

**A. Sampling of geometry parameters and lift coefficients**

The airfoil geometry is parameterized by the class shape transformation[27] (CST) method. The upper and lower sides are represented by seven base functions of CST. In total, 14 parameters are involved to generate an airfoil shape. Because not all combinations of these 14 parameters can produce practical airfoil geometry, there are some constraints on airfoil shapes from an engineering perspective. Two airfoil databases are prepared to illustrate the advantage of the proposed model:

- Database I is designed according to the scenario in optimization, and the $(t/c)_{max}$ is 0.095 for all airfoils, which is a typical relative thickness of the airfoil for a wide-body aircraft. Then, the output space sampling (OSS)[28] method is applied to sampling in the space of 14 geometry parameters. The purpose of sampling is to obtain airfoil geometries with abundant upper-surface pressure distribution, which is crucial for the off-design performance of a transonic airfoil. Meanwhile, other engineering constraints, including that the leading-edge radius is larger than 0.007 and the cruise point drag coefficient does not exceed 0.1, are employed to avoid impractical airfoils. A total of 1498 samples are obtained in the sampling process, 1400 are used to train the model, and the rest are used for testing.

- Database II is designed to test the model's prediction ability when using airfoils with highly diverse airfoil shapes, which is crucial when applying the model under consideration to an optimization process. Thus, a significantly more diverse testing airfoil database is designed, where 459 samples are obtained under $(t/c)_{max}$ values of 0.09, 0.10, 0.11, 0.12, and 0.13 by the OSS. The leading edge and drag coefficient constraints are the same as those in database I.

The airfoil shapes in the above two databases are depicted in Fig 2.

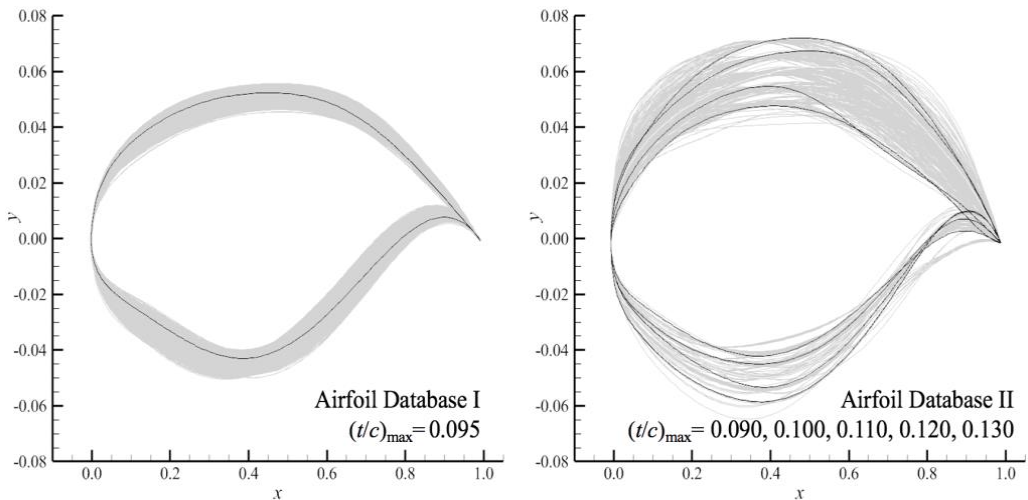

**Fig 2.** The airfoil geometries in Databases I and II

Several lift coefficients are computed by CFD based on the airfoil geometries of databases I and II. $C_L = 0.80$ is defined as the cruise point. The flowfield of training airfoils in Database I under four different $C_L$s around the cruise point are selected for training, while six other $C_L$s are selected for testing the model's prediction ability of the off-design flowfield both inside and outside the training range. For the airfoils in Database II, eight $C_L$s around the cruise point and inside the training range are used to test the model's generalization ability with different airfoils. All of the cases are computed by CFD with the fixed lift coefficient scheme; that is, the *AoA* is automatically adjusted to trace the required lift coefficient in the CFD iteration. The samples can be divided into groups based on their lift coefficient and thickness; the attributes of these groups are listed in Table 1.

**Table 1 The attributes of sample groups divided into different datasets**

| $(t/c)_{max}$ | Number of samples | $C_L$ | | | | | | | | | | |
|---|---|---|---|---|---|---|---|---|---|---|---|---|
| | | 0.60 | 0.64 | 0.68 | 0.72 | 0.76 | **0.80** | 0.84 | 0.88 | 0.92 | 0.96 | 1.00 |
| 0.090 | 93 | | ○ | ○ | ○ | ○ | | ○ | ○ | ○ | ○ | |
| 0.095 | 1400 | ◎ | ● | ◎ | ● | ◎ | | ◎ | ● | ◎ | ● | ◎ |
| | 98 | | | | | | | | | | | |
| 0.100 | 121 | | | | | | ☆ | | | | | |
| 0.110 | 90 | | ○ | ○ | ○ | ○ | | ○ | ○ | ○ | ○ | |
| 0.120 | 51 | | | | | | | | | | | |
| 0.130 | 6 | | | | | | | | | | | |
| | | | | | | (Training range of $C_L$) | | | | | | |

☆     Cruise point data (evaluated by CFD)
●     Training dataset
◎     Testing dataset of airfoils (Database I) with the same relative thickness
○     Testing dataset of airfoils (Database II) with different relative thicknesses

During the training and testing process, the *AoA* of each flowfield is used as an input value to the model to represent the freestream condition. The probability density distributions of the *AoA* for the datasets are checked and shown in Fig 3. The distributions nearby obey the Gaussian distribution, which suggests that the proposed data sampling method is practical.

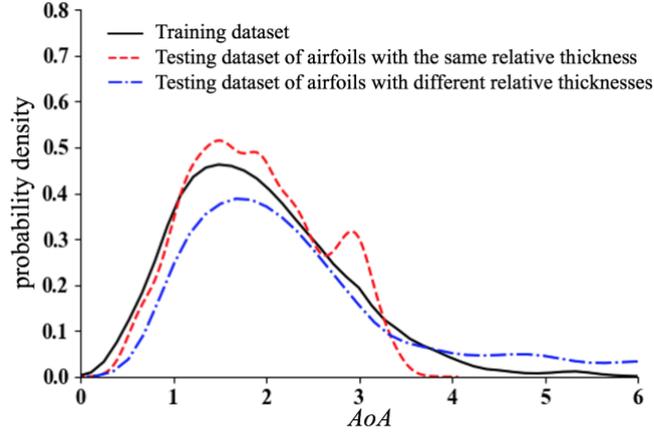

Fig 3. The density distribution of samples' *AoA* in the databases

## B. CFD methods

The structured C-type grid of an airfoil is automatically generated using an in-house built code, which solves an elliptic equation to ensure grid orthogonality. The grid is shown in Fig 4. The grid size is 381×81 in the circumferential direction (*i*-direction) and wall-normal direction (*j*-direction), respectively. The grid contains 300 cells on the airfoil surface. The far-field location is 80 chords away from the airfoil. The height of the first mesh layer is 2.7e-6 chord to fulfill the condition of $y^+ \approx 1$, which is the requirement of the turbulence model.

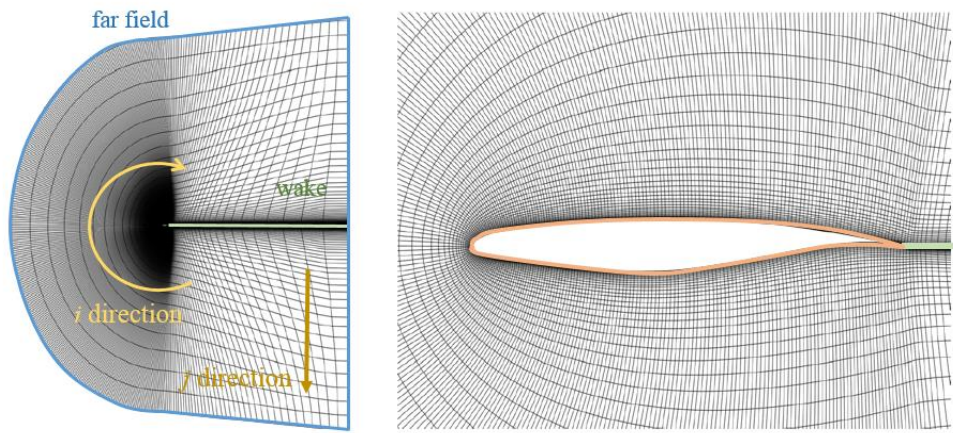

Fig 4. The C-type and its indexing scheme for airfoils

The flowfield is computed using the Reynolds Average Navier–Stokes (RANS) solver

CFL3D[29], which has been widely employed in engineering applications. The solver is based on the finite volume method. The MUSCL scheme, ROE scheme, and Gauss–Seidel algorithm are adopted for flow variable reconstruction, spatial discretization, and time advance, respectively. The shear stress transport (SST) model is adopted for turbulence modeling. These computational settings were used in our previous studies[16][23]. The pressure coefficient ($C_p$) distribution of a typical supercritical airfoil RAE2822 are computed with different grid size and compared with the experimental results**Error! Reference source not found.**, as shown in Fig 5. The three grids have 201, 301, and 401 grid points on the airfoil surface, respectively.

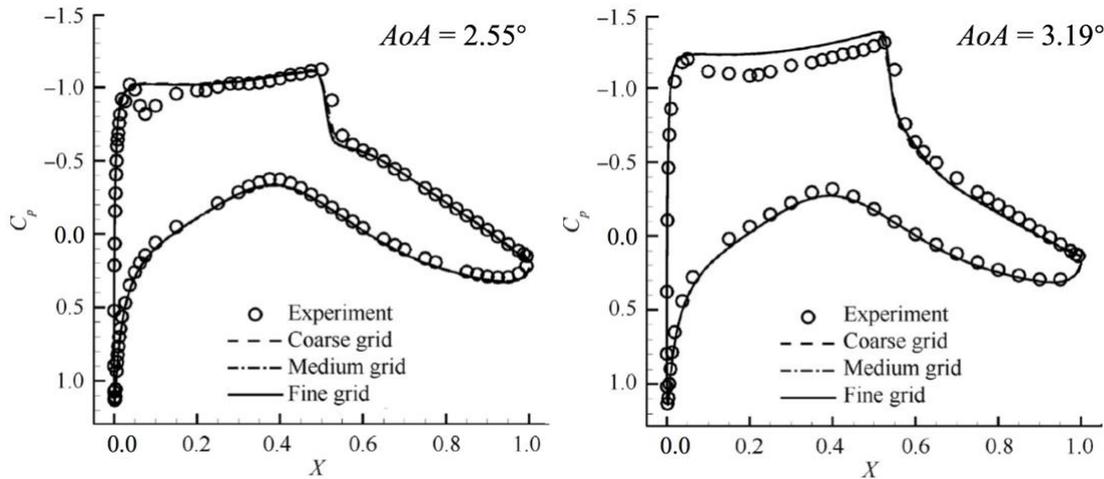

**Fig 5.** The pressure coefficient distribution of the RAE2822 airfoil under *Ma*=0.725, *Re*=6.5×10$^6$ and *AoA*=2.55° (left), *Ma*=0.73, *Re*=6.5×10$^6$ and *AoA*=3.19° (right)

Because the current study is based on a two-dimensional airfoil, four primary flow variables, pressure $p$, temperature $T$, velocity $u$, and velocity $v$, are chosen to represent a flowfield. Because each flow variable has a data channel and each mesh point $(i, j)$ represents a pixel of the flowfield, the structured CFD result can be reshaped to a matrix of 4×381×81 float numbers, as shown in Fig 6.

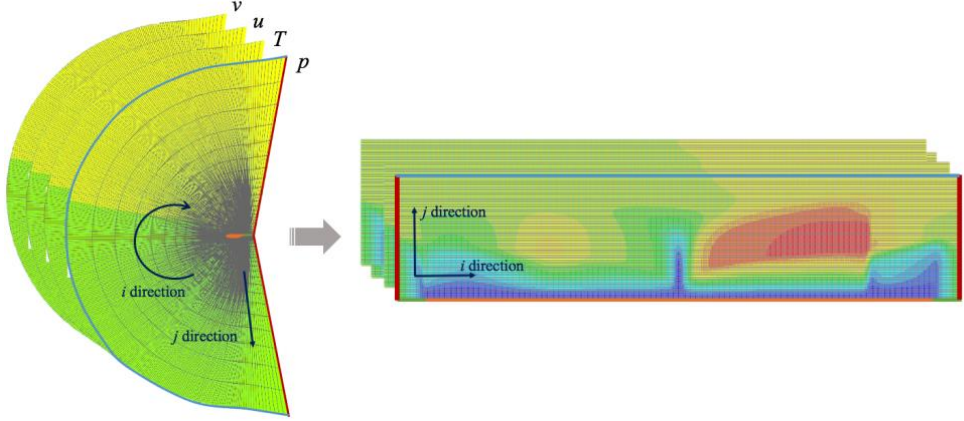

**Fig 6.**     The method for transforming a CFD result to a matrix

Each flow variable has its range and is quite different from others and thus must be normalized to improve the model's learning efficiency. However, the normalization methods using all the training data may fail because of outliers, so each flowfield is nondimensionalized based on the freestream parameters. The freestream pressure $\tilde{p}_\infty$, temperature $\tilde{T}_\infty$, and sound speed $\tilde{a}_\infty$ are used to nondimensionalize[29] the values in each flowfield, as shown in Eqn. (1). $\gamma$ is the specific heat ratio of the air. $R$ is the gas constant. The variables with tildes are dimensional, and the variables without tildes are the nondimensionalized variables used in the training and testing processes.

$$p = \frac{\tilde{p}}{\gamma \tilde{p}_\infty}, T = \frac{\tilde{T}}{\tilde{T}_\infty}, u = \frac{\tilde{u}}{\tilde{a}_\infty}, v = \frac{\tilde{v}}{\tilde{a}_\infty}, \tilde{a}_\infty = \gamma R \tilde{T} \qquad (1)$$

The values of all variables are transformed to the vicinity of 1 when using this normalization method; meanwhile, the relationships between variables are maintained.

## III. Off-design flowfield prediction model

In this section, the DNN framework for predicting off-design flowfields based on the reference cruise flowfield is proposed. The concept of the model is depicted in Fig 7.

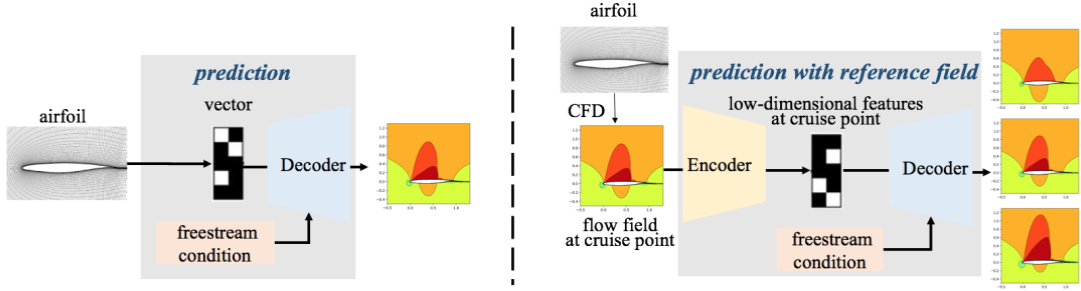

**Fig 7.    The previous multicondition flow prediction model (left) and the proposed model with a reference flowfield (right)**

In contrast to previous flow prediction models whose input is a vector or a distance map, the input of the new model is a reference flowfield that has the same structure as the output flowfield. To generate the series of flowfields from the reference, low-dimensional features first must be extracted from the reference flowfield and then concatenated with the freestream code to produce new flowfields. In the field of computer vision, the autoencoder (AE) is developed to extract features of images. As shown in Fig 8, an AE consists of two parts: the encoder compresses the image to low-dimensional latent codes, which contain the features of the image, while the decoder reconstructs the image from the codes. The objective of model training is to minimize the difference between the reconstructed image and the original image. A flowfield can be seen as an image, so models based on AE can be applied to extract the features of flowfields.

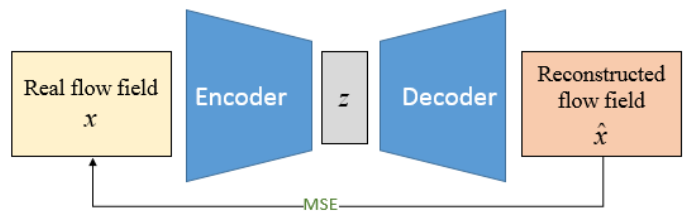

**Fig 8.    The framework of an autoencoder (AE)**

In the following section, the method of AE is revised first, and then, the theoretical basis of the new model is derived to predict multicondition flowfields with reference data.

Then, the derivation of physics-based loss terms is demonstrated.

## A. Variational Autoencoder

Based on the concept of an autoencoder, the variational autoencoder[31] (VAE) is developed from the perspective of Bayesian learning to gain better performance. Because the theoretical basis of the new model is VAE, a brief introduction to VAE is presented here.

Different from AE, the encoder and decoder in VAE are probabilistic, which means they learn posterior distributions rather than mapping relationships, as shown in Fig 9. Specifically, the encoder acts as a recognition model, which takes the real flowfield, $x$, as its input and produces an approximate distribution $q_\varphi(z \mid x)$ over the possible values of the latent code, $z$. The distribution is supposed as a Gaussian distribution with mean $\mu$ and standard deviation $\sigma$, which are the parameters ($\mu$, $\sigma$) that the encoder outputs. Meanwhile, the decoder acts as a generative model, which reproduces a distribution $p_\theta(x \mid z)$ over the possible values of the flowfield. However, the decoder actually outputs the reconstructed flowfield rather than a distribution of it. The reconstructed flowfield can be seen as the maximizing a posteriori estimation (MAP) of the distribution, as follows: $\hat{x} = \arg\max_x \mathrm{E}_{q_\phi(z|x)} p_\theta(x \mid z)$.

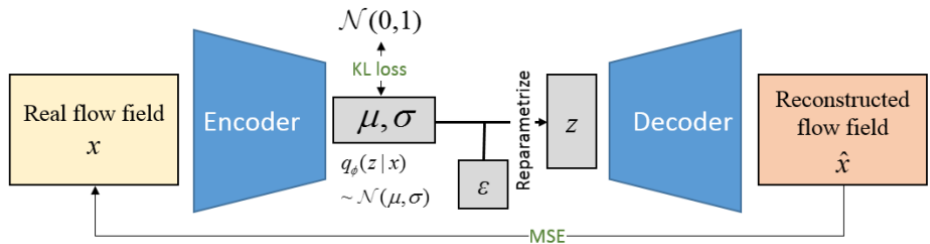

**Fig 9.** The framework of a variational autoencoder (VAE)

From a probabilistic perspective, the goal of model training is to obtain the two posterior distributions according to the sample set. With the variational approximation method[31] being applied, the encoder is derived to minimize the Kullback-Leibler divergence (*KL* divergence) between the reconstructed and the real distribution: $\arg\min_{\phi} KL(q_\phi(z|x) \| p_\theta(z|x))$, while the goal of the decoder is to maximize the possibility of generating a real sample: $\arg\max_{\theta} \mathrm{E}_{z \sim q_\phi(z|x)} p_\theta(x|z) = \arg\max_{\theta} p_\theta(x)$.

Those two goals can be combined by maximizing the lower bound on the evidence (ELBO)[31]:

$$ELBO_{\mathrm{VAE}} = \mathrm{E}_{z \sim q_\phi(z|x)} \log p_\theta(x|z) - KL(q_\phi(z|x) \| p_\theta(z)) \qquad (2)$$

Here, $p_\theta(z)$ stands for the prior distribution of *z* that is independent from *x* and is supposed to be a standard Gaussian distribution. Then, the Monte Carlo gradient estimator and the reparameterization trick are applied[31]. For each data point, a certain latent code *z* is randomly sampled from $q_\varphi(z|x)$ to produce a reconstructed flowfield through the decoder; then, the first term in Eqn. (2) can be substituted by the MSE distance between the reconstructed field, $\hat{x}$, and the real flowfield, *x*.

As recommended by Higgins[32], an adjustable hyperparameter beta is introduced to balance the independence constraints of the latent code with reconstruction accuracy. Thus, the loss function of VAE is:

$$\mathrm{L}_{\mathrm{VAE}}(\theta, \phi; x) = \frac{1}{2}\|\hat{x} - x\|^2 + \beta \cdot KL(q_\phi(z|x) \| p_\theta(z)) \qquad (3)$$

The connection with the autoencoder model is obvious since the first term expects a negative reconstruction error, while the second term acts as a regularization to avoid overfitting. From the perspective of flowfield order reduction, the latent code is the lower-dimensional representation of the flowfield.

## B. Model framework for predicting the multicondition flowfield with prior reference

The previous VAE model can generate a new flowfield by randomly sampling from the prior distribution of latent code, $p_\theta(z)$. In contrast, a multicondition prediction model is required to generate a flowfield under the constraint of the freestream condition. To fulfill this demand, the latent code, $z$, is divided into two parts, $z_c$ and $z_f$, as shown in Fig 10. The first part, $z_c$, represents the freestream condition, and the second part, $z_f$, is the features extracted from the flowfield at the cruise condition. The encoder inferences both $z_c$ and $z_f$, referred to as the recognition model $q(z_c, z_f | x)$, while the decoder produces a reconstructed flowfield from $z_c$ and $z_f$, referred to as the generative model $p(x | z_c, z_f)$.

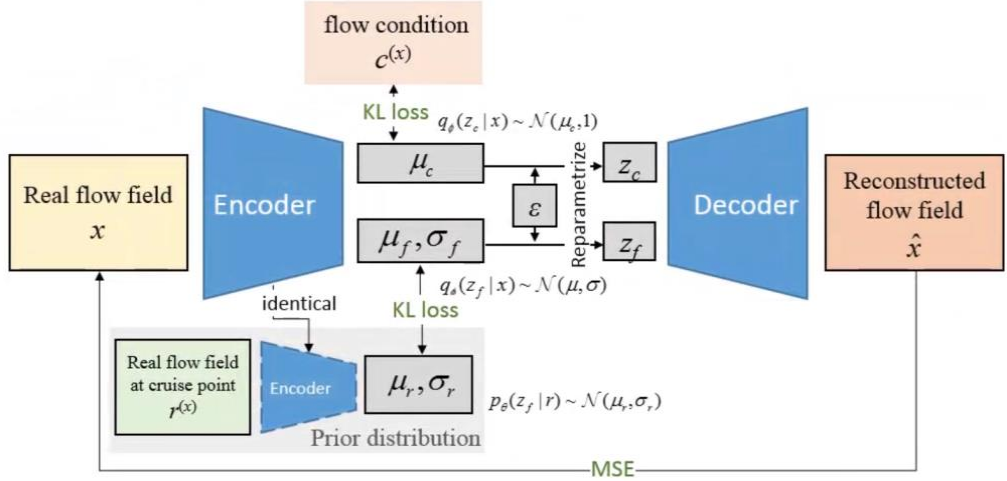

Fig 10. The prior-VAE (pVAE) framework for predicting multicondition flowfields

In a similar vein to VAE, the deduction of the training goal starts from minimizing the $KL$ divergence between the approximate posterior distribution $q_\phi(z_c, z_f | x)$ and the real distribution $p_\theta(z_c, z_f | x)$. The new $KL$ divergence is defined as:

$$KL\big(q_\phi(z_c, z_f | x) \| p_\theta(z_c, z_f | x)\big) = \mathrm{E}_{q_\phi(z_f, c | x)} \left[ \log \frac{q_\phi(z_c, z_f | x)}{p_\theta(z_c, z_f | x)} \right] \quad (4)$$

Considering the definition of the conditional probability, $q_\phi(z_c, z_f | x)$ can be expanded as $q_\phi(z_f | z_c, x) \cdot q_\phi(z_c | x)$; then, we have:

$$KL = \mathrm{E}_{q_\phi(z_f|x)} \log \frac{q_\phi(z_f | z_c, x)}{p_\theta(z_f | z_c, x)} + \mathrm{E}_{q_\phi(z_c|x)} \log \frac{q_\phi(z_c | x)}{p_\theta(z_c | x)} \quad (5)$$

The freestream condition of each flowfield $x$ is known. Therefore, $p_\theta(z_c | x)$ is a delta distribution. Supposing that $q_\phi(z_c | x)$ is a Gaussian distribution with a mean $\mu_c^{(x)}$ that is given by the encoder and the standard variance, $\sigma_c^{(x)}$, is set to 1, and let $c^{(x)}$ be the freestream condition corresponding to the input flowfield, $x$, then the second term in Eqn. (5) can be rewritten as follows, since $\left(\mu_c^{(x)} - c^{(x)}\right)^2$ is small:

$$\mathrm{E}_{q_\phi(z_c|x)} \log \frac{q_\phi(z_c | x)}{p_\theta(z_c | x)} = \frac{1}{\sqrt{2\pi}} \exp\left(-\frac{1}{2}\left(\mu_c^{(x)} - c^{(x)}\right)^2\right) \approx -\zeta \cdot \frac{1}{2}\left(\mu_c^{(x)} - c^{(x)}\right)^2 \quad (6)$$

Let $r^{(x)}$ be the flowfield at the cruise point of an airfoil corresponding to $x$, which is introduced into the first term in Eqn. (5):

$$\mathrm{E}_{q_\phi(z_c, z_f|x)} \log \frac{q_\phi(z_f | z_c, x)}{p_\theta(z_f | z_c, x)} = \mathrm{E}_{q_\phi(z_c, z_f|x)} \log \left[\frac{q_\phi(z_f | z_c, x)}{p_\theta(z_f | z_c, r)} \cdot \frac{p_\theta(z_f | z_c, r)}{p_\theta(z_f | z_c, x)}\right] \quad (7)$$

It is obvious that the flowfields of one airfoil under different freestream conditions refer to one $r^{(x)}$, so $z_f$ is only determined by $r$. Referring to the derivation of VAE and considering the independence of $c$ and $z_f$, the *ELBO* of prior VAE (pVAE) can be obtained:

$$ELBO_{\mathrm{pVAE}} = \mathrm{E}_{q_\phi(z_c, z_f|x)} \log p_\theta(x | z_f, z_c) - KL\left(q_\phi(z_f | x) \| p_\theta(z_f | r)\right) \\ - \zeta \cdot \frac{1}{2}\left(\mu_c^{(x)} - c^{(x)}\right)^2 \quad (8)$$

Compared with VAE's *ELBO*, the first term plays a similar role as that in Eqn. (2),

which is to reduce the reconstruction error. The last term is an additional term in pVAE, intended to introduce the freestream condition. The $p_\theta(z)$ in the second term in Eqn. (2) is substituted with $p_\theta(z_f | r)$ in Eqn. (8), which means the prior distribution of $z$ relies on the cruise flowfield, $r$. Again, the same techniques as VAE are involved to solve the problem, yielding the loss function of pVAE as follows:

$$L_{pVAE}(\theta, \phi; x) = \frac{1}{2}\|\hat{x} - x\|^2 + \zeta \cdot \frac{1}{2}\|\mu_c - c^{(x)}\|^2 + \beta \cdot KL\left(q_\phi(z_f | x) \| p_\theta(z_f | r^{(x)})\right) \quad (9)$$

Suppose $q_\phi(z_f | x)$ and $p_\theta(z_f | r^{(x)})$ obey the Gaussian distribution, with the parameters ($\mu_f$, $\sigma_f$) and ($\mu_r$, $\sigma_r$), respectively. $\beta$ and $\zeta$ in Eqn. (9) are hyperparameters, employed to ease the independence requirement.

According to research on the conditional variational autoencoder[33] (cVAE), another network should be established to predict the parameters ($\mu_r$, $\sigma_r$) of the prior distribution from the input, $r$. However, as demonstrated above, $r$ has the same structure as $x$. During the training process, the approximate distribution tends to the real distribution: $q_\varphi(z_f | r)$ →$p_\theta(z_f | r)$. Thus, the term $p_\theta(z_f | r)$ in *KL* divergence can be substituted by the distribution predicted by the same encoder, $q_\varphi$. A prior coupled autoencoding variational Bayes (PC-AEVB) algorithm is developed to iteratively obtain the prior distribution, $p_\theta(z_f | r)$, and posterior distribution, $q_\varphi(z_f | x)$, with the same encoder.

**Algorithm 1** prior coupled autoencoding variational Bayes algorithm
> Initialize ($\mu_r$, $\sigma_r$)
> for epoch = 1, …, $N_{epoch}$:
> {
>   for airfoil sample No. *i*:
>   {
>     for each Condition $c$ (including cruise point), we have the real flowfield $x$
>     {
>       $\mu_c$, ($\mu_f$, $\sigma_f$) = Encoder inference the posterior distribution with input $x$

>         $\hat{z}_c, \hat{z}_f$ = reparametrize from $N(\mu_c, 1)$ and $N(\mu_f, \sigma_f)$
>         $\hat{x}$ = Decoder generate reconstructed flowfield with $\hat{z}_c, \hat{z}_f$
>         $g$ = Gradients estimated with $\nabla_{\theta,\phi} L_{MFAE}(\theta, \phi; x)$
>         $\theta, \phi$ = Update parameters using gradients
>         }
>     ($\mu_r$, $\sigma_r$) = Update the prior with the posterior at cruise point ($\mu_f$, $\sigma_f$), which is calculated in the loop above
>     }
> }

### C. Physics guiding strategy for generalization improvement

Loss terms based on flow physics may provide more information to the model when training and help avoid overfitting when using a limited dataset, which may improve the model's generalization ability. In this paper, two loss terms are proposed.

The flowfield near the airfoil are expected to be more crucial than other parts of the flowfield, since the elements near the airfoil primarily determine the performance of the airfoil. The aerodynamic forces, including the lift coefficient, $C_L$, and drag coefficient, $C_D$, are calculated by integrating the pressure force and shear stress around the wall boundary; therefore, these coefficients are selected to guide the pVAE model to focus more on the flow near the wall by adding their prediction error to the loss function. Consequently, the loss term of aerodynamic forces is expected to guide the model to better predict the flowfields of new airfoils.

The predicted lift and drag coefficients obtained from the reconstructed flowfield are denoted as $\hat{C}_L$ and $\hat{C}_D$. The coefficients computed from CFD are $C_L$ and $C_D$. Then, the aerodynamic loss term is defined as:

$$L_{Aero} = \left( \frac{|\hat{C}_L - C_L|}{C_L} + \frac{|\hat{C}_D - C_D|}{C_D} \right) \tag{10}$$

Conservation of mass is another fundamental physical law that applies to fluid flow. Therefore, the mass flux is calculated for the reconstructed flowfield and is minimized by adding a corresponding term to the loss function of pVAE. The mass flux is calculated on the reconstructed flowfield cell by cell. In the framework of the finite volume solver, the flow variables are stored at the cell center. The flow variables are interpolated from the cell center to the grid edge to calculate the mass flux, as shown in Fig 11.

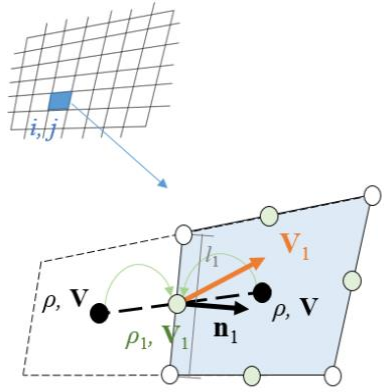

**Fig 11.  Mass flux calculation of a single cell in the 2D flowfield**

Then, the mass flux of the grid cell can be calculated as $\Phi_{m,ij} = \sum_{k=1}^{4} \rho_k (\mathbf{V}_k \cdot \mathbf{n}_k) l_k$, where the density, $\rho$, can be calculated based on the pressure and temperature of the cell. Finally, the mass residual is compared with the residual of the CFD result, and the positive deltas are counted for the loss function of the model:

$$L_{Mass} = \Phi_m = \max\left(0, \sum_{ij} \Phi_{m,ij} - \Phi_{m,ij}^{(real)}\right) \qquad (11)$$

## IV. Experimental Setup

### A. Neural Network Architecture

The conceptual framework of the proposed pVAE model is shown in Fig 10, above.
20

Considering the complex flow structure of flowfields, a 21-layer convolutional neural network based on ResNet[34] is chosen to set up the encoder and decoder for their favorable performance in deep feature extraction.

Fig 12 depicts the detailed architecture of the network. The encoder is composed of five groups, with each group containing two residual blocks (RB). One is a down-sampling residual block (RBDS), and the other is a basic RB. Two dense connected layers then give $\mu$ and $\sigma$, respectively. The number of latent variables is set to 12, including one latent variable representing the freestream condition. The decoder is similar to the encoder, while the down-sampling residual block is replaced with an up-sampling residual block (RBUS). At the end of the decoder, an additional convolutional layer is applied to finally generate the flowfield.

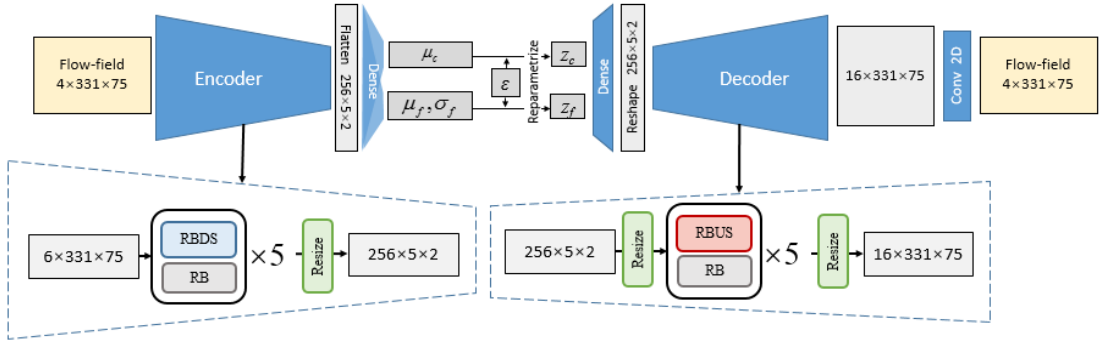

**Fig 12.**    The detailed architecture of the neural network

The architectures of the different residual blocks are depicted in Fig 13. A basic RB contains two routes. One is used for feature combination using convolutional layers, and the other is used for identity mapping. By adjusting the stride and padding parameters in the first convolution layers, the feature map size can be halved, which establishes the RB for down-sampling. Similarly, by adding a bilinear interpolation layer before the convolution layers, the feature map size can be multiplied, which creates the RB for up-sampling. In this study, LeakyReLU[35] with slope 0.2 is employed as the activation function. The network is implemented in the open-source package PyTorch.

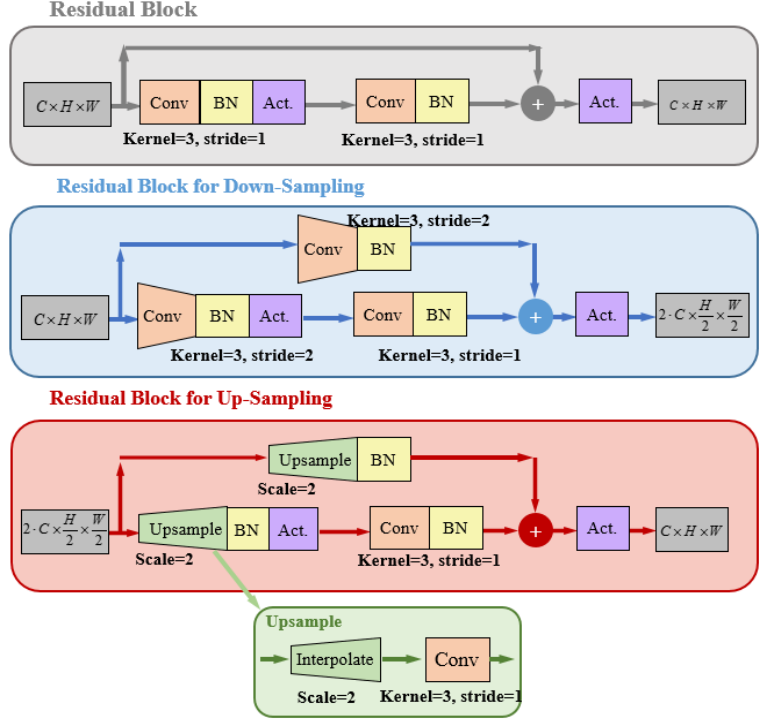

**Fig 13.** The architecture of residual blocks

(Gray: basic residual blocks, tin which he size of the output is the same as the input; Blue: down-sampling residual blocks, in which the size of output is halved and the channel number is doubled; Red: up-sampling residual blocks, in which the size is doubled and the channel number is halved)

In addition, a baseline model is set up to provide a benchmark for the proposed model. The baseline model has the same encoder and decoder as pVAE, but the input of the encoder is only the mesh of the airfoil. The architecture of the baseline model is shown in Fig 14.

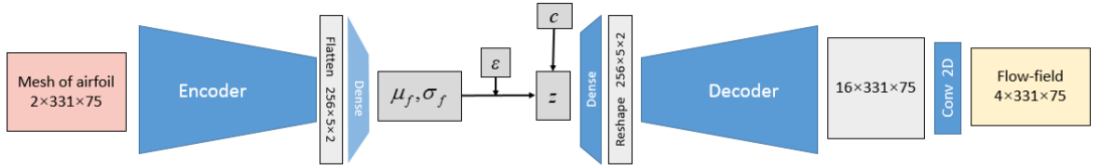

**Fig 14.** The architecture of the baseline model neural network

## B. Training process

The loss function of the model can be expressed as:

$$L_{pVAE} = L_{reconstruct} + \zeta \cdot L_{code} + \beta \cdot L_{prior} + \begin{cases} \lambda_{mass} \cdot L_{mass} \\ \lambda_{aero} \cdot L_{aero} \end{cases} \quad (12)$$

The expression for each term can be revised in Eqns. (9)–(11). Prior studies have been carried out to find the best weights for the loss terms; accordingly, $\beta, \zeta, \lambda_{mass}, \lambda_{aero}$ are set to 0.0001, 0.1, 0.0001, and 0.01 for optimal performance. To avoid instability, the weights of the physical loss terms are set to zero for the first 20 epochs.

The detailed training hyperparameters are selected as follows. A fixed batch size of 16 is applied, which can sufficiently balance the training efficiency and the performance of the trained model. The Adam algorithm[36] is selected as the optimizer to optimize the weights of the neural network. The warmup strategy[23] is employed to increase the learning rate from $5\times10^{-5}$ in the first 20 epochs so that instability at the beginning of the training process can be avoided. The ratio of the learning rate is written as:

$$\text{lr ratio} = \begin{cases} 1 + 0.5 \times \text{epoch}, & \text{epoch} < 20 \\ 10 \times 0.95^{\text{epoch}-20}, & \text{other} \end{cases}$$

Ten percent of the training samples are randomly selected as the validation dataset. The model is trained three times. Each of training run begins from a random initialization of weights in the model to avoid occasionality. All three runs converge after 300 epochs. The training and validation losses of one run are shown in Fig 15.

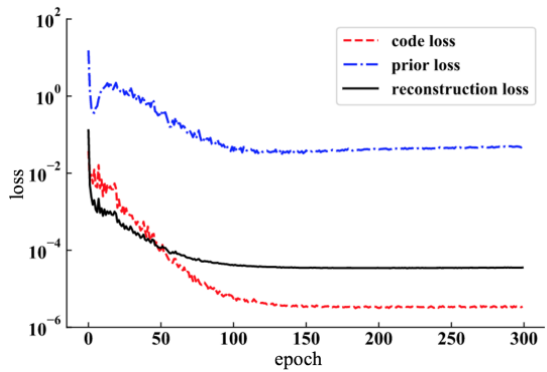

**Fig 15.   The training and validation losses for one run of the model**

## V. Results

The performance of the proposed model is assessed in this section. Its prediction accuracy and time consumption are compared with other methods to demonstrate the feasibility of the proposed model. The generalization ability of the model is also tested.

### A. Prediction ability for off-design flowfields at the testing $C_L$

Four models are trained with the flowfields of 1400 airfoils in Dataset I for comparison. The four models are the baseline model, pVAE model, pVAE model with a mass flow loss term, and pVAE model with an aerodynamic force loss term. Then, the models are used for predicting the flowfields of training airfoils under six testing $C_L$s.

The error of a flowfield is obtained by averaging the squared errors of every nondimensional flow variable and every grid, while the absolute errors of the aerodynamic force coefficients are used. The method used to calculate the errors for a sample is shown in Eqn. (13):

$$\Delta_{\text{flowfield}} = \left( \frac{1}{4N_{\text{grid}}} \sum_{\text{grid}} \left[ (\hat{p}-p)^2 + (\hat{T}-T)^2 + (\hat{u}-u)^2 + (\hat{v}-v)^2 \right] \right)^{\frac{1}{2}}$$
$$\Delta_{C_L} = \left| \hat{C}_L - C_L \right|, \Delta_{C_D} = \left| \hat{C}_D - C_D \right| \tag{13}$$

where the variables with hats are the results predicted by the models, and those without hats are the ground truth provided by CFD. Then, the errors are averaged by sample and by samples whose $C_L$s are within the range of training $C_L$s. The minimum averaged errors among the three runs are shown in Table 2.

Table 2 Prediction errors for test airfoils from Database I

| Test airfoil from Dataset I | Under all testing $C_L$ | | | Under testing $C_L$ within training range | | |
|---|---|---|---|---|---|---|
| Error of | flowfield | $C_L$ | $C_D$ | flowfield | $C_L$ | $C_D$ |
| baseline model | 0.0070 | 0.0164 | 0.0008 | 0.0048 | 0.0067 | 0.0005 |
| pVAE | 0.0078 | 0.0209 | 0.0013 | 0.0053 | 0.0102 | 0.0007 |
| pVAE with massflow residual | 0.0077 | 0.0162 | 0.0012 | 0.0055 | 0.0082 | 0.0007 |
| pVAE with aerodynamic loss | 0.0070 | 0.0148 | 0.0008 | 0.0048 | 0.0070 | 0.0003 |

Relatively good results are obtained for flowfield and aerodynamic force prediction for both the baseline model and pVAE model when the $C_L$ is in the training range, as shown in Table 2. The pVAE model does not illustrate its advantage over the baseline model. This is because the pVAE sacrifices a degree of accuracy to enhance its generalization ability. However, the airfoils in Database I are quite similar in that the generalization ability of pVAE is not useful here. The prediction accuracies of the baseline model and pVAE model decrease dramatically for $C_L$ beyond the training range. Both physical loss terms contribute to reducing the prediction error of the extrapolated freestream condition compared to the pVAE.

The CFD-simulated and model-predicted dimensionless pressure field of a randomly selected airfoil in Database I are depicted in Fig 16 to further demonstrate the prediction ability of the models. The dimensionless pressure profiles on the airfoil surfaces are also shown. The contour patterns and profiles show that flow structures such as shockwaves can be accurately reproduced, especially for freestream conditions inside the training range. For freestream conditions outside the training range, i.e., the first and last column in Fig 16, slight differences occur mainly at the shockwave region. In particular, a lower $C_L$ may lead to a double shock pattern, as shown in the first column of Fig 16, which is uncommon for flowfields under training conditions. In such cases, a single-shock pattern typically occurs.

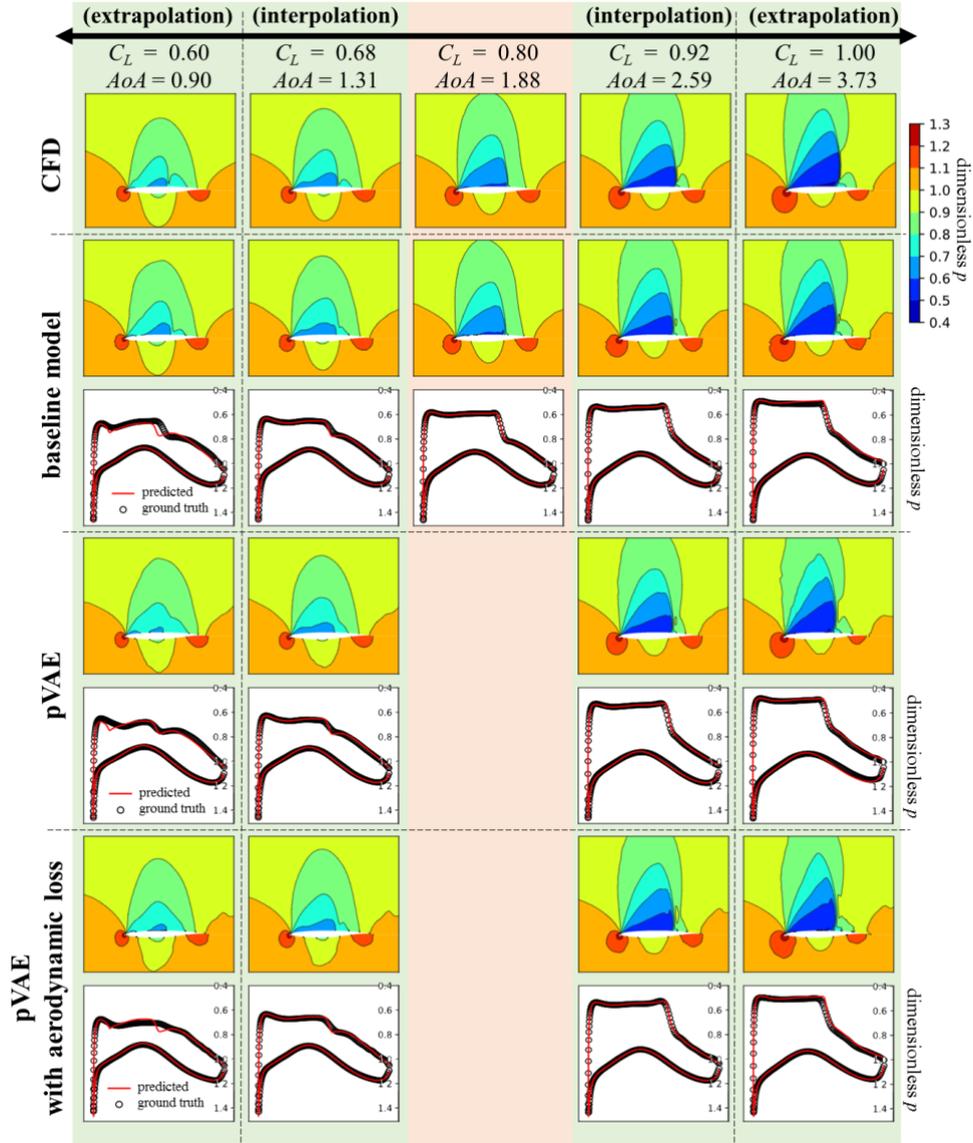

**Fig 16.  CFD-simulated and model-predicted pressure fields of a randomly selected airfoil from Database I under different freestream conditions**

(The third column depicts the flowfield under cruise conditions, i.e., $C_L$=0.80. The second and fourth columns show the flowfield under $C_L$ inside the training range. The first and last columns are outside of the training range.)

The aerodynamic curves of the same airfoil as that in Fig 16 are depicted in Fig 17, where both models with and without the reference cruise flowfield perform well under the training conditions (solid black dots), but the proposed model performs better beyond the training conditions (hollow black dots). Compared to the baseline model, the pVAE model can better capture the relationship among the flowfields of one airfoil, so the predicted curves are smoother and more precise.

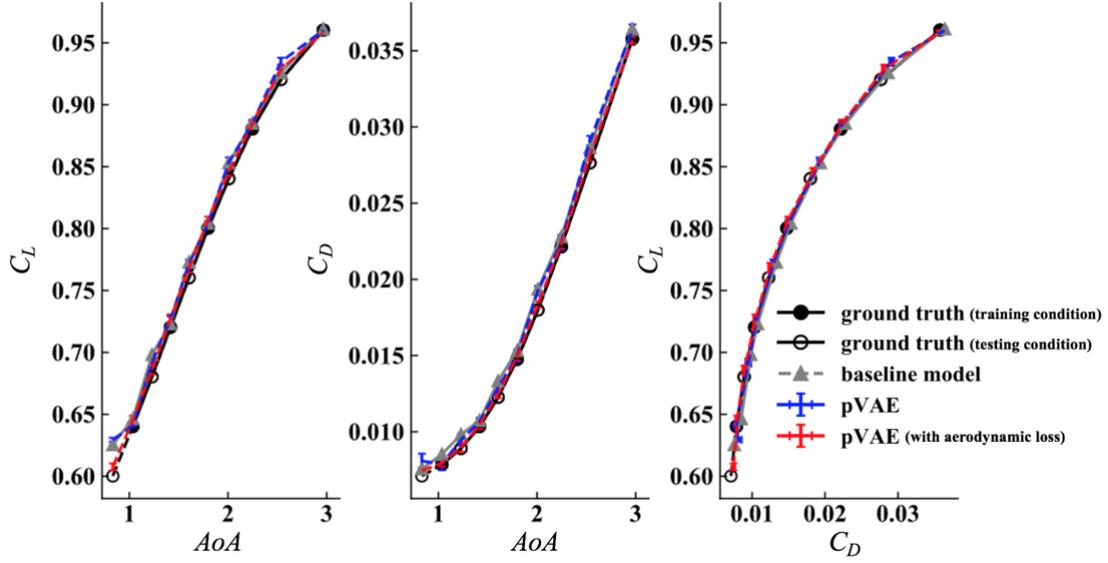

Fig 17. Aerodynamic curves of the airfoil in Fig 16

B. Generalization ability for more diverse airfoil geometries

The models are also tested with the airfoils in Database II that have different $(t/c)_{max}$ to compare their generalization ability in regard to more diverse airfoil shapes. The models are still trained with the 1400 airfoils contained in Database I. The errors are evaluated in the same way as in the section above with the results shown in Table 3. The prediction error of the flowfield is reduced by 30% by introducing the cruise flowfield compared to the baseline model. With physical-base loss terms, the error can be further reduced by 4%.

Table 3 Prediction errors for airfoils from Database II

| Test airfoil from Dataset II with $(t/c)_{max}$ = 0.090 and 0.010 | | | |
|---|---|---|---|
| *Error of* | flowfield | $C_L$ | $C_D$ |
| baseline | 0.0523 | 0.1952 | 0.0105 |
| pVAE | 0.0365 | 0.0810 | 0.0050 |
| pVAE with massflow residual | 0.0360 | 0.0768 | 0.0047 |
| pVAE with aerodynamic loss | 0.0346 | 0.0732 | 0.0032 |

The prediction errors, as shown in Fig 18, are averaged by airfoils with the same $(t/c)_{max}$

to make the comparison clear. The errors increase rapidly as the airfoil geometry gradually deviates from the training set. However, the pVAE performs much better than the baseline model because it has additional cruise flowfield information for the new airfoils. The physics guiding strategy is still effective with the new airfoil geometries to further reduce the reconstruction and prediction errors.

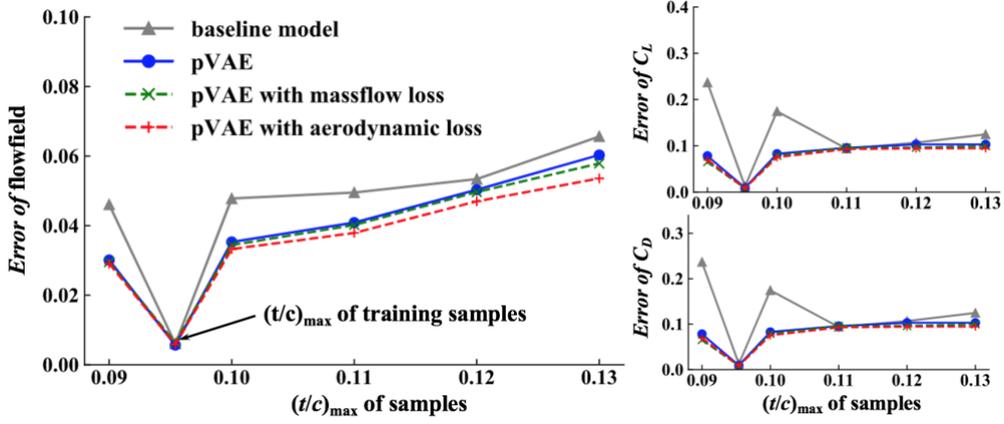

**Fig 18.** Averaged prediction errors of the models for test airfoils grouped by values of $(t/c)_{max}$

The prediction errors of the test airfoils with $(t/c)_{max}$=0.090 and 0.010 by model type are shown in Fig 19. The errors of the baseline model are more dispersed than those of the pVAE, which shows that the proposed pVAE model exhibits stable performance when dealing with unfamiliar airfoil geometries.

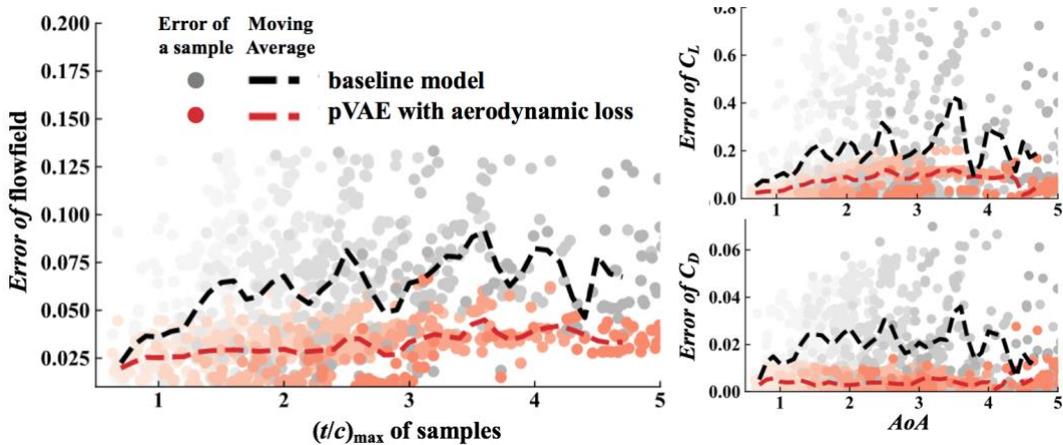

**Fig 19.** Prediction errors of the models with test airfoils

In Fig 20, the CFD-simulated and model-predicted dimensionless pressure fields of a randomly selected test airfoil with $(t/c)_{max}=0.090$ are depicted to further illustrate the advantages of pVAE. Intuitively, the prediction results of pVAE (the third row) are smoother and more consistent with the ground truth, while vibration and roughness occur in the flowfields predicted by the baseline model. The pressure profiles show that the pVAE model can better capture the flow features, such as the shockwave and suction peak. The double-shock pattern is still hard to predict at lower $AoA$s because of the lack of data in the training dataset.

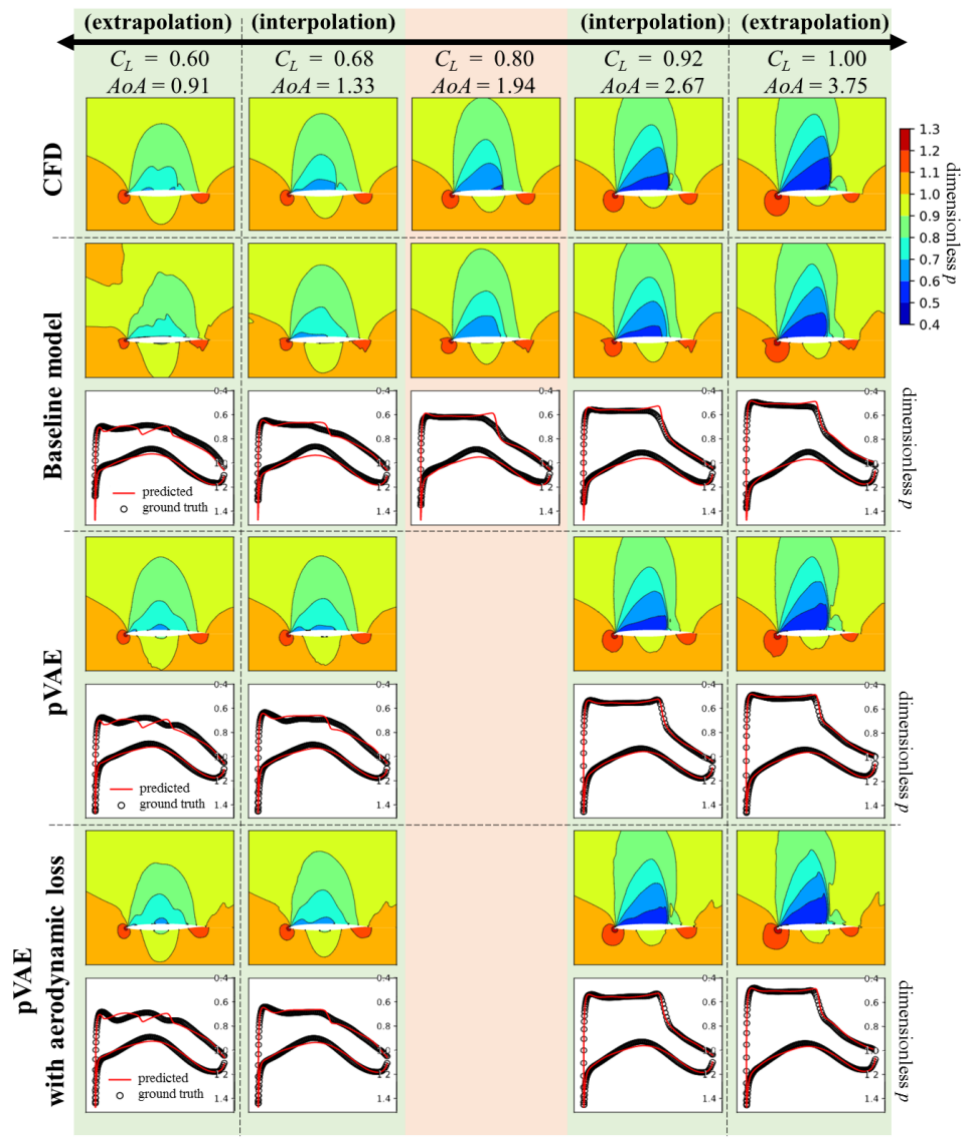

**Fig 20.** CFD-simulated and model-predicted pressure fields of a randomly selected airfoil from Database II under different freestream conditions

In addition, the pVAE model obviously predicts a much more precise pressure profile on the lower side of the airfoil. This is because the constraints during the process of sampling Database I lead to a similar pressure distribution on the lower side. However, when the $(t/c)_{max}$ of the airfoil changes, the pressure distribution also changes. The baseline model fails to predict this change because it lacks information regarding the variation of the lower side pressure distribution; in contrast, with the cruise flowfield provided to the pVAE model, it can better transfer the flowfield evolution learned on a limited airfoil dataset to a new airfoil geometry.

The above discussions have shown that pVAE models can improve the prediction accuracy for off-design point flowfields, especially when the airfoil shape is varied from the shapes used in training. The application of the ML method in airfoil shape optimization is a balance between time consumption and prediction error. Thus, the time consumption and drag prediction errors of different methods are listed in Table 4. The pVAE sufficiently balances the computation cost and fidelity requirement for both cruise and off-design points and has good application prospects in multipoint optimization.

Table 4 Time consumption and drag prediction errors of different methods

|  | Time consumption | | | Average drag coefficient error | |
| --- | --- | --- | --- | --- | --- |
|  | Data Preparation* | Training** | Prediction*, ** (7 off-design point) | design point | off-design point |
| CFD | / | / | 16 min | Zero | Zero |
| baseline | ~230 h | 5.3 h | 1.0 s | 0.0003 | 0.0006 |
| pVAE | ~230 h | 7.5 h | 2 min | Zero | 0.0003 |

\* CFD simulation is conducted on a single CPU of 2.00 GHz

\** ML model training and prediction are conducted on GPU, with a batch size of 16

# VI. Conclusion

Multipoint evaluation is crucial for airfoil shape optimization, but such activities are also time-consuming if every off-design point is evaluated by CFD. With the development of machine learning methods for use in in flowfield prediction, attempts have been made to apply such methods to optimization processes to reduce the associated time cost. However, the prediction fidelity of traditional ML methods cannot meet the optimization requirements; moreover, the model's generalization ability is also insufficient when facing airfoils whose geometries differ from those used in training. In this paper, a novel model framework, pVAE, is proposed to address these challenges. The contributions of our work can be concluded as follows:

1. A novel model framework based on VAE is proposed for the off-design point prediction of airfoils. Instead of predicting flowfields under all conditions, the cruise point is still evaluated by CFD, and the result is provided to the ML model as a reference for off-design point prediction. The performance of the model demonstrates that it sufficiently balances the time consumption and the different fidelity requirements of evaluation for cruise and off-design points, which makes this model feasible for use in an optimization process.

2. The effect on new airfoils of introducing a reference flowfield to the model's prediction and generalization ability are evaluated. The results show that the pVAE model can better capture the relationship between a series of flowfields and is able to transfer the knowledge learned on the limited dataset to new airfoils. Thus, the pVAE predicts a more precise flowfield and aerodynamic coefficients when the airfoil geometry deviates from the training geometries. The error is reduced by 30% when using the test airfoil database compared to that of the baseline model.

3. The errors of aerodynamic coefficients between the prediction and the ground truth are added to the model's loss function, which can further reduce the flowfield prediction error by 4%, and the prediction error is especially reduced for samples at $C_L$ that are beyond the training conditions.

Nevertheless, further investigations might be carried out to combine the proposed model with a proper optimization algorithm, where the assistance of our model for optimization can be illustrated and more useful results may be produced. It is hoped that our work will provide a new perspective for improving the model's generalization ability and a starting point for new tools for airfoil shape optimization.

**Acknowledgements:**

This work was supported by the National Natural Science Foundation of China under Grant Nos. 92052203, 91852108 and 11872230. The authors would like to thank Chongyang Yan and Yuqi Cheng for their inspiring comments on the work.